\documentclass[a4paper,fleqn,usenatbib]{mnras}

\usepackage[dvipsnames]{xcolor}

\usepackage[T1]{fontenc}
\usepackage{ae,aecompl}

\usepackage{graphicx}	
\usepackage{amsmath}	

\title[Globally magnetized discs]{A simple model of globally magnetized accretion discs}

\author[M. Begelman]{
Mitchell C. Begelman$^{1,2}$\thanks{E-mail: mitch@jila.colorado.edu}
\\
$^{1}$JILA, University of Colorado and National Institute of Standards and Technology, 440 UCB, Boulder, CO 80309-0440, USA \\
$^{2}$Department of Astrophysical and Planetary Sciences, 391 UCB, Boulder, CO 80309-0391, USA
}

\date{Accepted 2024 October 1. Received 2024 September 30; in original form 2024 February 23
}

\pubyear{2024}

\begin{document}
\label{firstpage}
\pagerange{\pageref{firstpage}--\pageref{lastpage}}
\maketitle

\begin{abstract}
We present an analytic, quasi-local model for accretion discs threaded by net, vertical magnetic flux. In a simple slab geometry and ignoring stochastic mean-field dynamo effects, we calculate the large-scale field resulting from the balance between kinematic field amplification and turbulent diffusion.  The ability of the disc to accumulate magnetic flux is sensitive to a single parameter dependent on the ratio of the vertical diffusion time to the Alfv\'en crossing time, and we show how the saturation levels of magnetorotational and other instabilities can govern disc structure and evolution.  Under wide-ranging conditions, inflow is governed by large-scale magnetic stresses rather than internal viscous stress.  We present models of such ``magnetically boosted'' discs and show that they lack a radiation pressure-dominated zone.  Our model can account for ``magnetically elevated'' discs as well as instances of midplane outflow and field reversals with height that have been seen in some global simulations. Using the time-dependent features of our model, we find that the incorporation of global transport effects into disc structure can lead to steady or episodic ``magnetically arrested discs'' (MADs) that maximize the concentration of magnetic flux in their central regions.   
\end{abstract}




\section{Introduction}

Magnetohydrodynamic (MHD) simulations of accretion discs, in both local (shearing-box) and global setups, yield drastically different outcomes depending on the presence or absence of a large-scale magnetic field, and its strength.  An organized vertical magnetic field ($B_z$) of sufficient strength, carrying a net magnetic flux that threads the disc, can have dramatic effects on the inflow speed and vertical disc structure.  Even for thin, gas pressure-supported discs, the $B_z B_\phi$ stress can dominate the extraction of angular momentum \citep{blandford82,ferreira93, ferreira95}, leading to much higher inflow speeds than produced by internal viscous stresses as in the model of \cite{shakura73}.  Moreover, such discs can become ``magnetically elevated" at sufficiently strong $B_z$, with a combination of turbulent and laminar magnetic pressure taking over from gas or radiation pressure as the main source of support. This effect thickens the disc and increases its inflow speed further \citep{bai13,salvesen16,zhu18,lancova19,mishra20}.  For a given accretion rate the disc density is correspondingly reduced, which can  significantly affect the disc's thermodynamics and radiative transport. 

Magnetic boosting of the inflow speed as well as magnetic elevation can be critical for understanding the dynamics, spectra and stability of accretion discs in a variety of astrophysical systems. Even before these computational trends had been established, the existence of magnetically supported discs had been proposed for phenomenological reasons \citep{pariev03,blaes06,begelman07,gaburov12,sadowski16,begelman17,dexter19}. Yet while magnetic elevation seems to be a robust feature of simulated flows, there is to date no clear explanation of how large-scale, organized magnetic field structures are created, and what factors determine their properties.

Even more extreme are so-called magnetically arrested discs (MADs) \citep{narayan03,igumenshchev2008}, which  have been studied intensively as a type of accretion flow that maximizes the magnetic flux threading a black hole, thus maximizing the potential to extract energy from the black hole's spin \citep{tchekhovskoy11}.  While the scaling between the accretion rate and trapped magnetic flux in a MAD was initially derived by equating the ram-pressure of free-fall accretion to the magnetic pressure of the trapped flux \citep{bisnovatyikogan74,bisnovatyikogan76,narayan03} --- arguing that the magnetic flux would ``arrest" the flow --- it has since been recognized that MAD models permit continuous accretion and that the accumulation of magnetic flux in a MAD can extend to much larger radii than the black hole horizon, in effect creating a large-scale magnetosphere \citep{mckinney12} co-existent with the accretion flow. 

Despite the potential for magnetically elevated and arrested discs to explain a wide range of accretion phenomena, there remains deep uncertainty about how to establish and maintain them in the first place.  A large body of work, beginning with \cite{vanballegooijen89} and \cite{lubow94}, suggests that under most conditions the flux distribution will evolve with time.  In particular, \cite{lubow94} highlighted the challenge faced by thin discs in accumulating and retaining flux, finding that the disc will advect and retain flux only if the poloidal field lines at the disc surface make an angle with the vertical smaller than $\sim H v_r / \eta $, where $H$ is the disc scale height, $v_r$ is the inflow speed inside the disc, and $\eta$ is the magnetic diffusivity in the disc interior, in most cases assumed to be due to turbulence.  For standard viscous disc models this critical angle is $\sim H/r \ll 1$, whereas constraints imposed by the external electromagnetic environment of the disc --- whether by matching the interior field to a magnotocentrifugal wind \citep{blandford82,konigl89,li95} or some other physically motivated field configuration \citep{lubow94,okuzumi14,takeuchi14} ---  tend to favor an angle of order unity. 

Various effects may help a disc to retain magnetic flux, including thickening of the disc due to magnetic pressure support, reduced diffusivity in the disc surface layers \citep{lovelace09,guilet12,guilet13}, and enhanced inflow speeds due to large-scale magnetic torques \citep{blandford82,ferreira93,konigl89,li95}.
In this paper, we focus on the latter effect, arguing that enhanced inflow speeds may be a natural feature of accretion discs under many conditions, even in the standard thin-disk limit in which the disc is supported by gas pressure and the magnetorotational instability (MRI) operates as usual.  

To make this case --- and explore its broader implications --- we construct and analyze a quasi-local, time-dependent, analytic model for thin discs threaded by a net vertical magnetic field.  The plan of the paper is as follows.  In \S\ref{sec:dynamo} we introduce the model (\S\ref{sec:dynamo1}) and solve for the magnetic field structure using boundary conditions (\S\ref{sec:boundary}) that connect the disc to its external environment.  We use these solutions to derive disc evolution equations (\S\ref{sec:discevol}), which show that magnetic stresses are at least as important for driving inflow and magnetic flux evolution as internal viscous stresses, and are often much more important.  We refer to these rapidly inflowing discs as ``magnetically boosted.''  We conclude the section with a discussion of the reduced levels of dissipation expected inside magnetized discs (\S\ref{sec:dissipation}), one consequence of which is the absence of any radiation pressure-dominated inner zone (\S\ref{sec:weaksteady}).

The analysis of \S\ref{sec:dynamo} reveals the critical importance of a parameter (labeled $q$) which essentially represents the ratio of the diffusion time across the disc scale height to the Alfv\'en crossing time.  Given $B_z$, this ratio mainly depends on the level of turbulence inside the disc, driven by magnetorotational and perhaps other instabilities.  In \S\ref{sec:MRI}, we discuss how this ratio might differ in weakly and strongly magnetized discs, while recognizing that the saturation levels of disc turbulence are among the least understood inputs for this or any similar model.  Nevertheless, the adoption of a specific closure relation allows us to construct steady-state models of magnetically boosted accretion discs in \S\ref{sec:steady}, for both weakly and strongly magnetized limits.  We provide a preliminary, local analysis of the stability of these models in an Appendix. 

In \S\ref{sec:astro}, we take the first steps toward placing our model in an astrophysical context, asking what the disc behavior will be if we impose a fixed $B_z$ and $\dot M$ at some outer radius.  To some extent, the outcome depends on how the gas is injected, but some intringuing --- and fairly dramatic --- trends emerge.  For example, at very high $\dot M$ (relative to $B_z^2$), only a thin, viscous disc is possible, whereas in the opposite limit the disc should evolve to a MAD.  Moreover, our analysis suggests that MAD states formed in this way may be naturally episodic, although this deduction is highly subject to the very poorly understood turbulence closure.     

We summarize our results in \S\ref{sec:conclusions} and expand on two further implications: the possibility of midplane outflow and field reversals within the disc, and the potential relevance of our model to episodic accretion phenomena such as X-ray nova outbursts and changing-look AGN. 

\section{Magnetized disc model}
\label{sec:dynamo}

Extensive literature exists on accretion disc dynamos, aimed at understanding both the turbulent and organized magnetic fields \citep[e.g.,][and references therein]{blackman15}.  Turbulent accretion disc dynamos, which have been the subject of numerous numerical and analytic studies  \citep{kato95, vishniac97, ogilvie03, gressel10, kapla11, gressel15, squire15, ebrahimi16, dhang20,mondal23}, can operate with or without the presence of net magnetic flux, and may well be important in magnetized discs. An alternate, plausible scenario --- which we adopt here --- is that a large-scale field is imposed on the disc, either at an outer boundary where matter is introduced into the disc or through diffusion from an ambient medium.  Sources of net magnetic flux might include the stellar magnetosphere or stellar wind in a mass-transfer binary, or a large-scale field threading the molecular clouds in a galactic nucleus harboring an AGN.  If a conserved net vertical flux threads the inner disc, mean-field dynamo growth
from turbulence is not essential to the model, which relies solely on advected flux and shear amplification to supply poloidal and toroidal fields. This is not a dynamo in the classic sense that all components of the large-scale field need to be maintained through a self-consistent cycle.  Instead, the (primarily vertical) field associated with the net flux is taken to be indestructible over global radial and temporal scales under consideration --- it can only be redistributed within the disc radially.  Other field components --- primarily radial and azimuthal --- evolve through the action of shear (the $\Omega$-effect) and inflow inside the disc, while affecting the near-disc environment through vertical diffusion and stress.
 
There is an extensive literature adopting this kinematic approach, much of it concerned with connecting the disc's large-scale field with magnetocentrifugal winds \citep{konigl89,li95,ferreira95,ferreira97}.  This paper falls in this camp, although we focus almost exclusively on the structure of the disc interior and consider the wind primarily as a sink of energy and angular momentum --- mainly through its Poynting flux.  An unusual feature of our approach is that we abandon the assumption of radial self-similarity, which is adopted in most treatments in the literature.  As our objective is to isolate the most robust features of magnetically elevated discs, we create a ``minimal" model that nevertheless admits physically interesting analytic solutions.  We ignore all stochastic mean-field effects except for a turbulent diffusivity.  While we calculate the radial inflow speed self-consistently, we assume Keplerian angular velocity and adopt a simple model for the vertical distribution of density and turbulent pressure.      

\subsection{Magnetized slab disc}
\label{sec:dynamo1}

We base our model on the induction equation with a scalar turbulent diffusivity  $\eta$,  
\begin{equation}
    \label{eq:dynamo1}
    {\partial{\bf B} \over\partial t} = - \nabla \times {\bf E} =  \nabla \times \left[ {\bf v} \times {\bf B} - \eta (\nabla \times {\bf B})\right] \ , 
\end{equation}
but for simplicity assume that the diffusivity only acts on derivatives with respect to $z$ (in cylindrical coordinates).  This assumption is generally consistent with a thin disc (scale height $H \ll r$), provided that $|B_r| \ga (H/r) B_z $ near the disk surface (where we take $B_z$ to be positive). At this point the source and level of turbulence responsible for $\eta$ are unspecified; in \S\ref{sec:MRI} we will discuss how it might be set quantitatively by the magnetorotational instability (MRI).

We assume that the disc can be represented by an axisymmetric, differentially rotating slab with density $\rho(r,t)$ and $\eta (r,t)$ independent of $z$ for $-H(r,t) < z < H(r,t)$, and zero elsewhere. The angular velocity $v_\phi(r)$ is assumed to be close to Keplerian and approximated as independent of height, while the radial velocity, $v_r$, can be a strong function of $z$. Since the vertical velocity $v_z \sim v_r (d H/d r) (z/r)$, we can ignore terms containing $v_z$ provided that $|d H/dr |\ll 1$.  

Given that the components of $\bf B$ are constrained by the condition
\begin{equation}
    \label{eq:divb}
\nabla \cdot {\bf B} = B_z' + {1\over r} {\partial \over \partial r} \left( r B_r\right)  = 0   
\end{equation}
(where partial derivatives with respect to $z$ are denoted by a prime), we only need two components of the simplified induction equation, e.g.
\begin{equation}
    \label{eq:Bphidot} 
    {\partial B_\phi \over\partial t} =   \eta B_\phi''- {3\over 2} {v_\phi B_r \over r} - {\partial \over \partial r} (v_r B_\phi)  
\end{equation}
\begin{equation}
    \label{eq:Bzdot}
    {\partial B_z \over\partial t} =  - {1\over r} {\partial \over \partial r} \left( r \eta B_r' + r v_r B_z\right) \ .
\end{equation}
The characteristic timescale for variations of $B_z$ is  
\begin{equation}
    \label{eq:tz}
 t_z \sim \min \left[{r\over v_r}, {B_z \over B_r }{r H \over \eta} \right] \ ,
\end{equation}
whereas for $B_\phi$ it is $t_\phi \sim H^2 / \eta \ll t_z $.  Therefore, it is reasonable to assume that $\partial B_\phi /\partial t \sim 0 $ over relaxation timescales for $B_z$.  We can also ignore the last term in equation (\ref{eq:Bphidot}), which corresponds to the evolution of $B_\phi$ due to radial flow, assumed to be sub-dominant compared to the $\Omega$-effect associated with rotational shear (represented by the penultimate term). Equation (\ref{eq:Bphidot}) thus takes the simple form 
\begin{equation}
    \label{eq:Bphidot2} 
    \eta B_\phi''= {3\over 2} {v_\phi B_r \over r}  \ .  
\end{equation}

Now assume that $B_z$ is approximately independent of $z$, i.e. $B_z (r,t)$.  This proves to be an excellent approximation for $H/r \ll 1$.   We define the magnetic flux interior to $r$, 
\begin{equation}
    \label{eq:Phidef} 
    \Phi (r,t) \equiv   \int^r_0  r dr B_z  \ ,  
\end{equation}
so that equation (\ref{eq:Bzdot}) can be integrated to yield
\begin{equation}
    \label{eq:Phidot}
     \dot\Phi  =  - \left( r \eta B_r' + r v_r B_z\right) \ ,
\end{equation}
where the dot denotes a partial derivative with respect to time.  The radial velocity is determined by the rate of angular momentum extraction, which is governed by a combination of large-scale magnetic torques and turbulent viscosity.  We approximate this by 
\begin{equation}
    \label{eq:vr}
     v_r  =   2 {B_\phi'B_z \over 4\pi \rho \Omega}  - {3 \over 2} {\rm Pr}_t {\eta \over r} \ .
\end{equation}
The final term on the right-hand side of equation (\ref{eq:vr}) represents the angular momentum loss due to internal viscous stresses within the disc, where ${\rm Pr}_t$ is the turbulent magnetic Prandtl number (the ratio of turbulent viscosity to magnetic diffusivity) and we have adopted the expression for a steady-state Keplerian disc for simplicity.  The first term on the right-hand side represents the divergence of the large-scale stress, where we have kept only the term associated with the vertical flux of angular momentum; here $\Omega$ is the Keplerian angular velocity.  In the applications below, the term corresponding to the radial flux is never larger than the viscous term and is typically smaller. 

Substituting for $v_r$ and exploiting the slab-disc assumption that $\rho$, $\eta$ and $B_z$ (and thus $\Phi$) are independent of $z$, we can integrate equation (\ref{eq:Phidot}) over $z$, using the fact that $B_r = B_\phi =0$ at $z=0$ (by symmetry) and rearranging terms to obtain
\begin{equation}
    \label{eq:Breq}
     B_r  =   - { 2 v_{{\rm A}z}^2 \over \eta\Omega} B_\phi + \left( {3 \over 2} {\rm Pr}_t B_z  - {\dot \Phi \over \eta } \right)   {z \over r}  \ ,
\end{equation}
where $v_{{\rm A}z} = (B_z^2 /4\pi\rho)^{1/2}$ is the Alfv\'en speed associated with the vertical field.  Inserting this expression into equation (\ref{eq:Bphidot2}) then gives an equation for $B_\phi(z)$:
\begin{equation}
    \label{eq:Bphieq}
     B_\phi'' + 3 \left( {v_{{\rm A}z}^2 \over \eta } \right)^2 B_\phi + {3\over 2} {\Omega\over\eta} \left( {\dot \Phi \over \eta } - {3 \over 2} {\rm Pr}_t B_z  \right) {z \over r} = 0  \ ,
\end{equation}
which can be integrated to give
\begin{equation}
    \label{eq:Bphieq2}
     B_\phi(z) = A \sin \left( {\sqrt{3}v_{{\rm A}z} \over \eta } z \right) -  {\eta \Omega \over 2 v_{{\rm A}z}^2} \left( {\dot \Phi \over \eta } - {3 \over 2} {\rm Pr}_t B_z  \right) {z \over r}  \ ,
\end{equation}
where $A$ is a constant of integration.  The corresponding expression for $B_r$ is 
\begin{equation}
    \label{eq:Breq2}
     B_r (z) = - { 2 v_{{\rm A}z}^2 \over \eta \Omega} A \sin \left( {\sqrt{3} v_{{\rm A}z} \over \eta } z \right)    \ .
\end{equation}

\subsection{Boundary conditions}
\label{sec:boundary}

In our simple slab geometry, we assume that the disc has a sharp boundary at $z = H$, where the density and diffusivity transition from finite $\rho, \eta $ to zero.  This is obviously an artificial assumption made for computational simplicity, but in fact it is physically reasonable.  Although there is no wind in our model, the magnetic field exterior to the disc carries both angular momentum and energy away from the disc.  Outside the disc, the field should adopt a force-free configuration quantitatively similar to disc-wind models in the limit that the Alfv\'en point along each streamline lies far outside its footpoint.  In the self-similar formulation of  \cite{blandford82}, this limit corresponds to $\lambda \rightarrow \infty$, $\kappa \rightarrow 0 $. 

The condition $\nabla\cdot {\bf B} = 0 $ implies that $B_z$ is continuous across the disc surface. If we assume that there are no surface currents so $\nabla \times {\bf B}$ is finite, then $B_\phi$ and $B_r$ must be continuous as well. 
Since the electromagnetic flux of angular momentum leaving the disc is proportional to $B_\phi B_z$, continuity of $B_\phi$ and $B_z$ ensures continuity of this flux.  In order to avoid singularities in $\dot{\bf B}$, the electric field components $E_r$ and $E_\phi$ must also be continuous.  Denoting the disc interior as region 1 and the exterior as region 2, these conditions imply 
\begin{equation}
    \label{eq:Ejump}
    \eta B_{\phi 1}' = - B_z \left(  v_{\phi 1} - v_{\phi 2} \right) \ ; \ \ \   \eta B_{r 1}' = - B_z \left(  v_{r 1} - v_{r 2} \right) \ ,
\end{equation}
respectively.  Given the continuity of ${\bf B}$, these conditions ensure that the Poynting flux of energy leaving the disc is continuous.

The two additional conditions needed to specify a model are motivated by assumptions about the ``wind zone" outside the disc.  For one of the conditions, we adopt $B_\phi'(H)=0$. In our approximation scheme (neglecting the radial magnetic torque), $B_\phi'$ must vanish just outside the disc surface (where $\rho$ is vanishingly small) to avoid large  radial velocities (cf.~equation [\ref{eq:vr}]).  The condition $B_\phi'(H)=0$ therefore amounts to assuming that $B_\phi'$ is continuous across the boundary.  We note that \cite{ferreira93, ferreira95} demanded a similar condition in their mass-loaded wind models, arguing that the divergence of the angular momentum flux has to switch sign in order to transition from extracting angular momentum from the disc to feeding it to the wind. From equation (\ref{eq:Ejump}), we see that in our approximation scheme $B_\phi'(H)=0$ implies that $v_\phi$ is continuous across the boundary, but the same cannot be true for $v_r$.  

For the final condition, we adopt a parameter describing the tilt of the poloidal field at the disc surface,  
\begin{equation}
    \label{eq:xidef}
     \xi  \equiv {B_r(H) \over B_z }     \  .
\end{equation}
We expect $\xi$ to be determined by the structure of the magnetic field external to the disc, e.g. necessary to maintain force-free equilibrium near the disc surface or match to a wind solution at large $r$ \citep[e.g.,][]{blandford82,konigl89,li95,lubow94,okuzumi14,takeuchi14}.  Since $B_z$ is taken to be positive, $\xi > 0 \ (< 0)$ corresponds to poloidal field lines bending outward (inward) above the disc. We will generally assume that the field lines bend outward.

Applying these conditions to equations (\ref{eq:Bphieq2}) and (\ref{eq:Breq2}), we find \begin{equation}
    \label{eq:Acond}
     A = - \xi {\eta \Omega \over 2 v_{{\rm A}z}^2} {B_z \over \sin q}  \ ,
\end{equation}
where
\begin{equation}
    \label{eq:qdef}
     q  \equiv {\sqrt{3} v_{{\rm A}z} H \over \eta }     \ ,
\end{equation}
and 
\begin{equation}
    \label{eq:Ejump4}
     \left( {3 \over 2} {\rm Pr}_t  - {\dot \Phi \over \eta B_z}  \right) {H \over r} =   \xi   q  \cot q     \ .
\end{equation}

\subsection{Disc evolution equations}
\label{sec:discevol}

Using the definition of the enclosed flux, we can write equation (\ref{eq:Ejump4}) in the form  
\begin{equation}
    \label{eq:fluxeq}
    {\partial\Phi \over \partial t} +  \left(   \xi  q  \cot q      - {3 \over 2} {\rm Pr}_t {H\over r}  \right) {\eta \over H} {\partial\Phi \over \partial r}=  0   \ ,
\end{equation}
which shows that the flux is simply advected at a speed
\begin{equation}
    \label{eq:fluxadv}
    v_B =   \left(  \xi  q  \cot q   - {3 \over 2} {\rm Pr}_t {H\over r}  \right) {\eta \over H}     \ .
\end{equation}
This differs from the mean radial speed of the accretion flow, which is given by 
\begin{equation}
\label{eq:accretionspeed}
    \bar v = {1 \over H} \int_0^H v_r dz = v_B - \xi {\eta\over H} =  \left[  \xi  \left( q  \cot q - 1 \right)  - {3 \over 2} {\rm Pr}_t {H\over r}  \right] {\eta \over H}     \ .
\end{equation}
The mass conservation equation is then given by
\begin{equation}
    \label{eq:masseq}
    {\partial  \over \partial t} \left( \rho H \right) +  {1\over r} {\partial \over\partial r} \left(  \rho r H \bar v \right)  =  0   \ .
\end{equation}

The model is subject to two important dynamical constraints.  First, we assume that the vertical electromagnetic torque carries angular momentum away from the disc, rather than towards it; this implies $B_\phi (H) / B_z < 0$ or, equivalently, $\xi (q \cot q - 1) < 0$.  For the fiducial case where the field lines bend outward ($\xi > 0$), the requirement is $q \cot q < 1$, which is automatically satisfied for $q < \pi$.  Note that this condition on the angular momentum flux is stronger than the condition that the disc be accreting, i.e., that $\bar v < 0$, because of the additional contribution of the internal viscous torque.

Second, the pressure near the midplane must exceed the pinching force exerted by $B_\phi$ and $B_r$.  As emphasized  by \cite{ferreira93}, both of these pressure forces are directed toward the midplane at small $z$, and thus contribute to confining the disc vertically.  The scale height of the disc is determined by some internal pressure, $P_d$, which could be the usual gas or radiation pressure in the disc interior or the same turbulent pressure that determines $\eta$. The requirement that gravity ultimately confine the disc,  $P_d \sim \rho H^2 \Omega^2$, imposes the self-consistency condition that  $B_\phi^2 + B_r^2 < 8\pi \rho H^2 \Omega^2$ at all $z < H$.  A necessary condition,  $B_\phi^2(H) < 8\pi \rho H^2 \Omega^2$, gives $\xi^2 (q \cot q - 1)^2 < 8 q^2 / 3$. For $\xi \la 1$, this condition permits all values of $0 < q < \pi/2$ and $\xi$-dependent bands at higher $q$. 
Additional constraints may restrict the parameter space further. 

A steady state is attained for $v_B = 0$ and an accretion rate $\dot M \propto \rho r H \bar v$ independent of radius.  Applying the first condition with equation (\ref{eq:fluxadv}), we obtain a mean accretion speed
\begin{equation}
\label{eq:accretionspeed2}
    \bar v = - { \sqrt{3} \xi \over q }  v_{{\rm A}z}  = - \xi {\eta \over H}    \ .
\end{equation}
The magnetic flux is stationary when $\xi q \cot q = (3/2){\rm Pr}_t (H/r)$. Since we expect $q \sim O(1)$ under the action of MRI in a standard thin accretion disc with a weak magnetic field (\S \ref{sec:MRI}), we recover the condition obtained by \cite{lubow94} for a thin disc to retain poloidal field, i.e., the field needs to be very weakly inclined with respect to the vertical, $\xi \sim  H/r$.  However, our model exhibits a novel feature if $q$ approaches $\pi /2$: in this case, $q \cot q \ll 1 $ and flux can be retained even if $\xi \sim O(1)$, as is required in most large-scale disc-wind models \citep[e.g.,][]{blandford82,konigl89,li95,lubow94,okuzumi14,takeuchi14}. The reason for this is twofold.  First, the inward advection speed increases due to large-scale magnetic torques.  Second, the vertical gradient of $B_r$ decreases near the disc surface, which reduces the outward diffusion.  Under these conditions the accretion is driven almost entirely by large-scale magnetic stresses.  We refer to such discs as ``magnetically boosted."

For $q > \pi/2$, $q \cot q$ becomes negative and we see from equation (\ref{eq:fluxadv}) that magnetic diffusion actually helps to trap the flux and draw it inward.  This is because the poloidal flux surfaces become convex-outward, and thus exert an inward tension force near the disc surface.    

\subsection{Dissipation}
\label{sec:dissipation}

As pointed out by earlier works \citep{blandford82,ferreira93,ferreira95}, discs accreting due to large-scale magnetic stresses are much less radiatively efficient than standard thin accretion discs.  Indeed, only a fraction $\sim H/r$ of the liberated gravitational binding energy is dissipated inside the disc.  Here we calculate the dissipation rate per unit surface area for our simple model.  

Energy is dissipated in our model through both viscosity and magnetic diffusion.  The viscous dissipation rate per unit area is simply given by
\begin{equation}
\label{eq:Fnu}
    F_\nu = {3\over 2} {\rm Pr_t} \eta \rho H \Omega^2    \ .
\end{equation}
To calculate the dissipation due to magnetic diffusion, we evaluate
\begin{equation}
\label{eq:Feta}
    F_\eta = {\eta \over 4\pi} \int_0^H (\nabla \times {\bf B})^2 dz =  {\eta \over 4\pi} \int_0^H  \left[ \left( B_\phi'\right)^2 + \left( B_r' \right)^2 \right] dz \ ,
\end{equation}
which yields
\begin{equation}
\begin{aligned}
\label{eq:Feta3}
    F_\eta = {\xi^2 q^2 \eta B_z^2 \over 4\pi H }   \Biggl[ g^2 \Biggl(  {2 \cos^2 q +1 \over \sin^2 q} &  - 3 {\cot q \over q} \Biggr)  \\    & + \left( \csc^2 q + {\cot q\over q} \right) \Biggr] \ ,
\end{aligned}
\end{equation}
where the term dependent on the parameter
\begin{equation}
\label{eq:gdef}
    g \equiv {\eta \Omega \over 2 v_{{\rm A}z} } = {\sqrt{3} \over 2 q}  {H \Omega \over v_{{\rm A}z} } 
\end{equation}
expresses the dissipation of $B_\phi$ and the other term refers to $B_r$.  In discs supported vertically by non-magnetic pressure, as in standard thin discs (with $q \sim O(1)$), $g \gg 1$ and the dissipation due to $B_\phi $ dominates. In this limit, with $\xi \sim O(1)$, the magnetic  and viscous dissipation rates are comparable.  In a steady state, we have
\begin{equation}
\label{eq:Feta2}
    F_\eta \approx  {3\over 4} \xi^2  \eta \rho H \Omega^2    \ ,
\end{equation}

We can compare these rates to the local release of gravitational binding energy due to accretion,
\begin{equation}
\label{eq:Facc}
    F_{\rm acc} = {1\over 2} \rho r H \bar v \Omega^2 = {\xi \over 2 } \eta \rho r \Omega^2   = {1 \over 8\pi} {GM \dot M \over r^3 } \ ,
\end{equation}
where $\dot M = 4\pi \rho r H |\bar v|$ is the accretion rate and we have used the accretion speed in the steady-state limit.  As noted by \cite{ferreira93}, the ratio of dissipation to the local release of binding energy is $\sim \xi^{-1} (H/r)$, i.e., small whenever the inflow speed is strongly boosted by magnetic torques. 

\section{MRI-driven  Turbulence}
\label{sec:MRI}

We have seen that the evolution of our simple slab disc model depends critically on the strength of turbulence expressed through the parameter $q$, which represents the ratio of $v_{{\rm A}z} H$ to the turbulent diffusivity $\eta$ or, in physical terms, the ratio of the diffusion time to the Alfv\'en crossing time.  While the complete dependence of the model on $q$ is complex (and probably reflects the model's limitations), for $q \ga 1$ we see that increasing $q$ (weaker turbulence) leads to stronger advection and retention of magnetic flux, through both faster inflow speeds (due to magnetic extraction of angular momentum) and weaker magnetic diffusion.

Moreover, these behaviors can be very sensitive to the level of turbulence when $q \approx \pi / 2$.  While the numerical values at which the behaviors set in are undoubtedly an artifact of the simplified model, the qualitative trends are likely robust and can be understood in terms of the slightly different behaviors of mass advection and magnetic flux advection in a diffusive disc, and their delicate balance. 

In this section, we discuss how the value of $q$ may be set by the saturation of MRI in weakly and moderately magnetized discs.  We note that other instabilities could also contribute to the turbulence, possibly reducing $q$ below the values based on MRI alone.

We first consider the saturated level of MRI in a weakly magnetized disc. Both local \citep[shearing box:][]{hawley95,bai13,salvesen16}   and global \citep{mishra20} MRI simulations find that the viscous $\alpha$-parameter \citep{shakura73} exhibits a scaling 
\begin{equation}
    \alpha \propto \beta_z^{-1/2}.
\end{equation}
between $\beta_z \approx  10^5$ and $\beta_z \approx 10$, where the parameter $\beta_z$ is defined in terms of an imposed vertical magnetic field $B_z$ and mid-plane gas pressure through $\beta_z = 8\pi P_g/B_{z0}^2$.\footnote{This correlation is distinct from the well-known result $\alpha \propto \beta^{-1}$, where in this case $\beta$ is calculated using the magnetic energy density associated with the fully developed turbulence \citep{blackman08}}.  If ${\rm Pr_t}$ is a constant, this implies that $\eta \propto v_{{\rm A}z} H $ and $q$ is independent of $\beta_z$.  Setting $\alpha \sim 10 \beta_z^{-1/2}$ \citep{salvesen16}, this gives $q \sim 0.25 {\rm Pr_t}$, which falls short of the value $q \approx \pi/2$ needed for a steady state unless Pr$_{\rm t}$ is considerably larger than one.  

However, the simulations also indicate that the MRI-driven turbulent energy density approaches and exceeds the gas pressure when the vertical magnetic field is still quite weak, at $\beta_z \la 10^2$.  The enhanced turbulent pressure will thicken the disc \citep{scepi24}, but also --- according to our disc model --- drive the Alfv\'en speed associated with the organized toroidal field, $v_{{\rm A}\phi}$, above the sound speed.  In this limit, the growth rate of MRI is suppressed \citep{blaes94}.  Quasi-linear arguments suggest that the saturated level of MRI-driven turbulence is proportional to the instability growth rate on scales $\sim H$ \citep{begelman23}, in which case the amplification of $B_\phi$ as $\beta_z$ decreases could drive $q$ towards or even beyond $\pi /2$, leading to a steady state (if $q \la \pi/2$) or the rapid accumulation of flux (if $q > \pi/2$). We will adopt the {\it ansatz} that for weakly magnetized discs, there is a critical $\beta_z$, $\beta_{\rm crit} \sim 10^2$, at which $v_B = 0 $ and a steady state can be maintained for both mass flux and magnetic flux.            

The situation is murkier for moderately to strongly magnetized discs, for which numerical studies of turbulent saturation levels are scarce.  Although linear growth rates for MRI are diminished, the reversal of the large-scale $B_\phi$ (and $B_r$) components across the midplane could drive tearing instabilities that sustain turbulence near the midplane \citep{begelman23}. The energy source for the turbulence is directly the reconnection of magnetic energy in $B_\phi$ (and $B_r$), and indirectly the rotational shear energy that creates the toroidal field.  Equating the dissipation rate $F_\eta$ (equation \ref{eq:Feta2}) with the ``universal'' dissipation rate of collisionless reconnection, $F_{\rm rec} \approx 0.1 v_{{\rm A}\phi} B_\phi^2/8\pi$ \citep[e.g.][]{yamada10}, and assuming $q \approx \pi/2$ and $H \Omega \approx v_{{\rm A}\phi}$, we obtain $v_{{\rm A}\phi} \approx 17 \xi^2  v_{{\rm A}z}$. For simplicity, we will adopt $ v_{{\rm A}\phi} = 10  v_{{\rm A}z}$ when analyzing steady state disc models with $v_{{\rm A}\phi} \gg c_s$.

A quasi-local analysis of MRI-like modes \citep{das2018} shows an increase of growth rates at $ v_{{\rm A}\phi} \ga (c_s v_K)^{1/2}$, with rates approaching $\Omega$ as $ v_{{\rm A}\phi} \rightarrow v_K$.  These very magnetized discs are probably geometrically thick.  Since the approximation $H \ll r$ is central to our simplified analysis, we will not consider this case in detail.

\section{Steady State Discs}
\label{sec:steady}

We now combine our disc model with the simple turbulence closures proposed in \S\ref{sec:MRI}, to calculate the structure of discs that are globally in a steady state.  For simplicity, we assume $\xi =1$ and $H/r \ll 1$, implying $\cos q \sim O(H/r)$. We can therefore approximate
\begin{equation}
\label{eq:etasteady}
    \eta = {2 \sqrt{3}\over \pi}  v_{{A}z} H     \ .
\end{equation}
We consider separately the weakly and strongly magnetized cases.
 
\subsection{Weakly magnetized discs}
\label{sec:weaksteady}

Adopting the closure relation for this case, we take $\beta_z = \beta_{\rm crit} \equiv 10^2 \beta_2$, implying 
\begin{equation}
\label{eq:vazcs}
    v_{{A}z} =  0.14 \beta_2^{-1/2} c_s      \ .
\end{equation}
This is also very close to the mean accretion speed $|\bar v|$, from equation (\ref{eq:accretionspeed2}).  We assume that the disc is supported vertically by gas pressure, implying $H = c_s/\Omega$.  Interestingly, the reduced level of dissipation inside the disk implies that a magnetically boosted, steady state disc cannot be supported by radiation pressure, as can be seen by the following argument.  Vertical support of a disc by radiation pressure requires a local flux $F_d (r) = (H/r) GMc /\kappa r^2$, where $\kappa$ is the opacity.  But we have seen in \S\ref{sec:dissipation} that the actual disc flux is roughly a fraction $H/r$ of the locally liberated binding energy, or  $\sim (H/r) GM \dot M / 8\pi r^3$ (cf.~equation \ref{eq:Facc}).  Comparing the two fluxes, we find that maintaining radiation pressure support requires $\dot M$ to be a specific function of radius, $\dot M \sim 8\pi cr/\kappa$. This condition clearly cannot be satisfied at all $r$ in a conservative accretion flow, with constant $\dot M$.\footnote{Expressed as $r \sim \kappa \dot M / 8 \pi c$, this is the condition for the ``trapping radius" \citep{begelman78}, where advection and diffusion of radiation are comparable.}   

The accretion rate can be normalized to the Eddington value using the electron scattering opacity, $\kappa_{\rm es}$:
\begin{equation}
\label{eq:mdot}
    \dot m \equiv  {0.1 \dot M c^2 \over L_E} = {0.1 \dot M \kappa_{es} \over 4\pi r_g c} = 0.015 \beta_2^{-1/2}\tau x \left({c_s \over c}\right) \left({\kappa_{\rm es} \over \kappa }\right)   \ ,
\end{equation}
where $L_E$ is the Eddington limit, $r_g = GM/c^2$, $\tau = \rho \kappa H$ is the disk optical depth, and $x= r/r_g$. We have included a radiative efficiency factor of 0.1 in the definition of $\dot m$ to facilitate comparison with standard accretion disc treatments such as \cite{shakura73}. 

The dissipation is the sum of contributions from magnetic diffusivity and viscosity, and is given (for $\xi = {\rm Pr_t} = 1$) by
\begin{equation}
\label{eq:Fd}
    F_d = F_\eta + F_\nu =   {9\over 4} \eta \rho H\Omega^2 = 0.35 \beta_2^{-1/2}\rho c_s^3  \ .
\end{equation}
This is balanced by the radiative flux, which we assume is diffusive ($\tau \gg 1$) and thermalized, implying
\begin{equation}
\label{eq:Frad}
    F_{\rm rad} = {a c T^4 \over 3 \tau} =  1.4 \times 10^{47} \tau^{-1}  \left({c_s \over c}\right)^8  \ {\rm erg \ cm^{-2} \ s^{-1} } \ .
\end{equation}

To express $F_d$ in terms of $\tau$, we need to know the dominant opacity.  We first consider Kramers opacity, $\kappa \approx \kappa_K = 6.6\times 10^{22} \rho T^{-7/2}$ [cgs], which normally dominates in the outer regions of thin discs.  Radiative equilibrium fixes a relationship between the sound speed and optical depth, 
\begin{equation}
\label{eq:taucs}
    {c_s \over c} =  1.4 \times 10^{-4} \beta_2^{-1/4} m^{-1/4} x^{-3/8} \tau^{3/4}  \ ,
\end{equation}
where $m \equiv M/M_\odot$.  We then use equation (\ref{eq:mdot}) to calculate various parameters characterizing the disc's radial structure, e.g.
\begin{equation}
\label{eq:tauKramers}
    \tau =  77 \beta_2^{5/17} \dot m^{4/17} m^{3/17} x^{1/34}   \ ,
\end{equation}
\begin{equation}
\label{eq:TKramers} 
    T =  8.3 \times 10^7 \beta_2^{-1/17} \dot m^{6/17} m^{-4/17} x^{-16/17} \ {\rm K} \ ,
\end{equation}
\begin{equation}
\label{eq:HKramers}
    {H \over r} =  3.6 \times 10^{-3} \beta_2^{-1/34} \dot m^{3/17} m^{-2/17} x^{1/34}  \ ,
\end{equation}
where we have taken $\kappa_{\rm es} = 0.34 \ {\rm cm^2 \ g^{-1}}$ and emphasize that $T$ is the temperature in the disc interior.  These scalings are qualitatively and quantitatively similar to a standard disc model in the Kramers opacity-dominated regime \citep{shakura73}, as is the boundary within which electron scattering begins to dominate over Kramers opacity, $\kappa_{\rm es} / \kappa = 1$,
\begin{equation}
\label{eq:esKramersweak}
    x_K =  3.4 \times 10^3 \beta_2^{-9/23} \dot m^{20/23} m^{-2/23}   \ .
\end{equation}  
 
We next consider the inner disc region, $x < x_K$, where the opacity is dominated by electron scattering.  We therefore take $\kappa = \kappa_{\rm es}$ and repeat the analysis above, obtaining
\begin{equation}
\label{eq:taucses}
    {c_s \over c} =  3.3 \times 10^{-4} \beta_2^{-1/12} m^{-1/6} x^{-1/4} \tau^{1/3}  \ ,
\end{equation}
\begin{equation}
\label{eq:taues}
    \tau =  9.6 \times 10^3 \beta_2^{3/8} \dot m^{3/4} m^{1/8} x^{-9/16}   \ ,
\end{equation}
\begin{equation}
\label{eq:Tes} 
    T =  3.2 \times 10^8 \beta_2^{1/12} \dot m^{1/2} m^{-1/4} x^{-7/8} \ {\rm K} \ ,
\end{equation}
\begin{equation}
\label{eq:Hes}
    {H \over r} =  6.9 \times 10^{-3} \beta_2^{1/24} \dot m^{1/4} m^{-1/8} x^{1/16}  \ .
\end{equation}
These scalings (and others that can be derived from them) resemble the standard accretion disc scalings in the scattering-dominated regime.  In both opacity limits, however, the systematic deviations of the results from standard disc models embody the lower radiative efficiencies of magnetized discs and the suppression of radiation pressure as a major source of vertical support, when $\dot m <1$.  

\subsection{Strongly magnetized ``magnetically elevated'' discs}
\label{sec:strongsteady}

As noted in \S\ref{sec:MRI}, in the strongly magnetized case we assume $H = v_{{\rm A}\phi}/\Omega$ and adopt the closure $v_{{\rm A}z} = 0.1 v_{{\rm A}\phi}$.  This implies that the accretion mass flux can be written
\begin{equation}
\label{eq:mdotstrong}
    \dot M  =  4\pi \rho r H |\bar v| = 10 {r \over \Omega} B_z^2   \ ,
\end{equation}
i.e, as a pure function of $B_z (r)$ and radial factors.  Conversely, the vertical magnetic field in a strongly magnetized steady-state disc has a unique radial dependence, 
\begin{equation}
\label{eq:mdotstrongB}
B_z = \left( {\dot M GM  \over 10 r^{5/2} } \right)^{1/2} \propto r^{-5/4} \ .
\end{equation}
  The corresponding magnetic flux enclosed within $r$ is given by 
\begin{equation}
\label{eq:fluxstrong}
    2\pi \Phi (r)  =  {8\pi \over 3 \sqrt{10}} (\dot M v_K r^2 )^{1/2} \propto r^{3/4}  \ ,
\end{equation}  
which shares the scaling of magnetically arrested discs (MADs), although with a coefficient somewhat smaller than typically found in simulations \citep{tchekhovskoy11,mckinney12}. 

Since gas pressure is unimportant for vertical support, there is no unique solution for the scale height but rather a tradeoff between $H$ and $\rho$.  We can write $\dot M = 0.4 \pi \rho H^2 v_K$ and normalize to the Eddington limit as before:  
\begin{equation}
\label{eq:mdotstrong2}
    \dot m \equiv  {0.1 \dot M c^2 \over L_E} =  10^{-3}  \tau h_{-1} x^{1/2} \left(\ {\kappa_{\rm es}\over \kappa} \right)    \ ,
\end{equation}
where we have parametrized $H/r = 0.1 h_{-1}$.  Assuming thermalized radiation and balancing the dissipated flux against radiative losses, we obtain the same expression for the temperature as in the scattering-dominated, weakly magnetized case (equation \ref{eq:Tes}). However, because the strongly magnetized disc is thicker and less dense,  the Kramers opacity is lower and only dominates the total opacity outside 
\begin{equation}
\label{eq:esKramersstrong}
    x_K =  5.1 \times 10^4 \dot m^{24/50} h_{-1}^{32/25} m^{2/25}   \ ,
\end{equation}
i.e., at much larger radii than in the weakly magnetized case (equation \ref{eq:esKramersweak}), for $h_{-1}\sim O(1)$.  Because Kramers opacity is depressed, thermalization may fail in the disc interior, especially at small $\dot m$.  Estimating the effective optical depth for thermalization, we find   
 \begin{equation}
\label{eq:esKramersstrong2}
    \tau^* = (\tau_K \tau_{\rm es})^{1/2} = 0.21 \dot m^{5/8} h_{-1}^{-2} m^{-1/16} x^{9/32}   \ .
\end{equation}
When $\tau^* < 1$ we can expect the disc to be hotter, with a spectrum formed by Comptonized bremsstrahlung.  These conditions may also be favorable to the thermal decoupling of electrons and ions to form a two-temperature plasma, a case we do not consider further here.
 
\section{Astrophysical Context}
\label{sec:astro}

A premise of this paper is that accretion discs could be supplied with net magnetic flux from their surroundings.  We have argued that the ability of a disk to concentrate flux by advecting it inward is sensitive to its level of magnetization (e.g., through the plasma $\beta$-parameter).  In this section we address the global question of whether a disc forced into a highly magnetized state at large radii will then evolve toward a similar state at smaller radii.  Given the unique radial dependence of the organized vertical magnetic field, $B_z \propto r^{-5/4}$, required for steady accretion in a strongly magnetized disc, this question is clearly crucial for assessing the ability of discs to  enter the MAD state.

First, however, we address the related question: when can a disc {\it avoid} being driven into a highly magnetized state, or even being affected by a weak but inevitable $B_z$ in the environment? In other words, when is the Shakura-Sunyaev solution a reasonable approximation?

Since these are global issues, we will need to assume boundary or initial conditions in order to address them.  A comprehensive treatment of plausible conditions is beyond the scope of this exploratory study, and we instead focus on one simple but plausible case.  We suppose that matter is injected at a rate $\dot M_0$, spread over some radial interval $\Delta r_0$ around an outer radius $r_0$, and that the disc is bathed in a uniform vertical magnetic field $B_0$.  Physically, such a situation could represent the region around the circularization radius in a Roche-lobe-overflow binary, where the field is initially embedded in the stellar wind of the mass-transferring companion.  Alternatively, it could represent the outer region of an AGN where the disc is fed from interstellar gas with a net background magnetization.  

\subsection{Shakura-Sunyaev--like discs}
\label{sec:SSlike}

 For $\beta_z > \beta_{\rm crit}$, we assume that $\xi \la H/r$ due to inefficient flux advection, and therefore that inflow is driven by the usual Shakura-Sunyaev viscous stress.  We therefore take $v = \alpha (H/r)^2 v_K$ with $\alpha = 10 \beta_z^{-1/2}$ (cf.~\S\ref{sec:MRI}).  The accretion rate can then be written
 \begin{equation}
\label{eq:mdotSS}
    \dot M = 5 \beta_{\rm crit}^{1/2} \left( {\beta_z \over \beta_{\rm crit}} \right)^{1/2} {H\over r }  {r \over \Omega} B_z^2 
    \ .
\end{equation}
This yields two interesting lower limits on accretion rates compatible with a Shakura-Sunyaev disc in the presence of a background magnetic field.  For
\begin{equation}
\label{eq:mdotSS2}
    \dot M < 0.5 \beta_2^{1/2} \left({H\over r} \right)_{-2} {r \over \Omega} B_z^2
    \ ,
\end{equation}
$\beta_z < \beta_{\rm crit}$,
where we have once again expressed $\beta_{\rm crit}$ in units of $10^2$ and now normalize $H/r$ to 0.01. Moreover, for  
\begin{equation}
\label{eq:mdotSS3}
    \dot M < 0.05  \left({H\over r} \right)_{-2} {r \over \Omega} B_z^2
    \ ,
\end{equation}
$\beta_z < 1$ and the standard vertical-field MRI is quenched.  Note that both of these limits are well below the characteristic strongly magnetized accretion rate given by equation (\ref{eq:mdotstrong}).

If $\dot M$ is so low that either of the conditions (\ref{eq:mdotSS2}) or (\ref{eq:mdotSS3}) is satisfied, then the standard Shakura-Sunyaev mode should not be self-consistent because magnetic stresses will take over both the vertical support and angular momentum transport. For accretion rates between the two limiting values, we would expect vigorous turbulence leading to a magnetically supported disc with a magnetically boosted inflow speed, both of which would exacerbate the departure from viscous disk behavior.  Paradoxically, in the more extreme limit of condition (\ref{eq:mdotSS3}), we face the opposite situation that the level of turbulence might drop significantly as instability is quenched by the strong vertical field; in this case, the disc might  effectively stop accreting.   

We have argued elsewhere \citep{begelman22} that MRI cannot always be quenched under the condition $\beta_z < 1$, since even a small toroidal field $B_\phi$ still supports the rapid growth of MRI and presumably its evolution into turbulence.  But without an imposed $B_\phi$, its presence depends on the initial action of a dynamo.  A disc evolving from high $\beta_z$ toward lower values will already contain an organized toroidal field that can perhaps sustain itself, but the same cannot be guaranteed at an outer boundary where the disc is set up in a state with very weak to nonexistent MRI in the first place.   

We therefore suggest that a system with matter being injected at a rate satisfying condition (\ref{eq:mdotSS3}) simply builds up its column density in place, until $\beta_z$ rises to the point that MRI takes hold and the system can accrete at a rate much higher than $\dot M_0$.  This would lead to short duty cycles of high accretion interspersed with long periods of quiescence, much like those found in X-ray novae.  We will address in a separate publication the applicability of our model to XRB quiescence and outbursts.

Finally, we note that conditions (\ref{eq:mdotSS2}) and (\ref{eq:mdotSS3}) are most stringent at large radii, since $H/r$ is typically a weak function of radius and we do not expect $B_z^2$ to increase inward as steeply as $\Omega / r$. 

\subsection{Formation of MAD-like states}
\label{sec:MADlike}

We next consider the case where disc matter is injected at the outer boundary in a highly magnetized state.  The fiducial accretion rate for these conditions is obtained by evaluating equation (\ref{eq:mdotstrong}) at the outer radius, 
\begin{equation}
\label{eq:mdotmag}
\dot M_{\rm mag} = 10 {r_0\over \Omega(r_0)} B_0^2 \ .
\end{equation}
To analyze disc evolution, we use the time-dependent equations (\ref{eq:fluxeq}) and (\ref{eq:masseq}) and make several approximations appropriate to the strongly magnetized limit.  Dropping subscripts so that $B_z \rightarrow B$ and $v_{{\rm A}z} \rightarrow v \propto B \rho^{-1/2}$, we set $\bar v = - a v$ and $v_B = - b v $, where $a$ and $b$ are positive constants with $a > b$.  This mimics the regime where inward field advection is occurring, although at a slower rate than mass advection.  Renormalizing time to $\tau \equiv b t$, setting $\chi \equiv b/a$, and noting that $H/r \propto v/v_K$, we obtain       
\begin{equation}
    \label{eq:fluxeq5}
   \dot B - (v'B + vB') - {vB\over r} = 0  \ ,
\end{equation}
\begin{equation}
    \label{eq:masseq5}
    \chi \left( 2\dot B - B {\dot v\over v} \right) - 2 v B' - {5\over 2} {vB\over r} = 0  \ ,
\end{equation}
where a prime and dot indicate $\partial / \partial r$ and $\partial/\partial\tau$, respectively.  

With $B_0 =$ {\it const.}, $v(r_0)$ becomes a proxy for the outer disc density or, more germanely, the column density $\rho H \propto \rho^{1/2}$.  If $\dot M_0 > \dot M_{\rm mag}$, then the column density increases linearly with time since the accretion rate cannot balance mass injection; therefore $v(r_0) \propto t^{-1}$.  The solutions of equations (\ref{eq:fluxeq5}) and (\ref{eq:masseq5}) quickly converge to a MAD-like radial scaling $B \propto r^{-5/4}$, with $v \propto t^{-1} r^{1/4}$, corresponding to $\rho\propto t^2 r^{-3}$. Formally, this solution exhibits a steady inward accumulation of magnetic flux, as distinct from the ``true'' steady-state solution discussed in \S\ref{sec:strongsteady}, which has $v_B = 0$.  However, this steady flux advection solution cannot persist, since the disc scale height satisfies $H/r \propto v r^{1/2} \propto t^{-1} r^{3/4}$.  The strong magnetization assumption fails when $v \la 0.1 c_s$, and assuming  $c_s / v_K \sim 10^{-3} - 10^{-2}$ as is typically found from thermal models for the outer disc, this implies that the strongly magnetized limit fails both at small radii for arbitrary times and at all radii out to $r_0$ at sufficiently late times.  We conclude that for $\dot M_0 > \dot M_{\rm mag}$, only the Shakura-Sunyaev thin disk solution provides a viable steady  state. 

Now consider the opposite case, $\dot M_0 < \dot M_{\rm mag}$.  Here gas is advected away from the injection zone faster than it can be replenished, so the column density decreases with time and $v$ increases.  Using equation (\ref{eq:vr}) and assuming magnetically boosted inflow, we can write a steady-state condition, $\dot M_0 = (B_\phi / B_z)_0 \dot M_{\rm mag}$, indicating that $B_\phi$ must be depressed below our earlier estimate, $B_\phi \sim 10 B_z$, in this limit.  To see that this must be the case, we note that $v(r_0) \sim B_0/\rho^{1/2}$ increasing with time would eventually exceed $v_K$, blowing the disc apart if $B_\phi$ were not suppressed.  We can guess at the properties of a steady-state solution if we assume that the disc is supported by the toroidal field with $H/r \sim 1$, implying $B_\phi \sim v_K \rho^{1/2}$.  The formula for the magnetically boosted accretion rate, $\dot M \sim r B_\phi B_z / \Omega \sim r^2 \rho^{1/2} B_z$, then gives one relation between $B_z$ and $\rho$, while the flux advection condition, $r v_r B_z \sim \dot M B_z /(\rho r) \sim$ {\it const.}, gives a second.  Combining these, we obtain $\rho \propto r^{-2}$, $B_z \propto r^{-1}$, $B_\phi \propto r^{-3/2}$.  In this picture, $B_\phi$ ``catches up'' to $\sim 10 B_z$ at a radius $r_1 \sim (\dot M_0/\dot M_{\rm mag} )^{1/2} r_0$ and converges to the standard highly magnetized solution (\ref{eq:mdotstrong}), which has $B_z \propto r^{-5/4}$ and $v \propto r^{1/4}$, corresponding to $\rho\propto r^{-3}$. 

Like the solution with $\dot M_0 > \dot M_{\rm mag}$, this steady flux advection solution also cannot persist to arbitrarily small radii, since the disc scale height satisfies $H/r \propto v r^{1/2} \propto r^{3/4}$.  As before, we would expect the strong magnetization assumption to fail when $v \la 0.1 c_s$, but unlike the case with high $\dot M_0$, the low-$\dot M_0$ state cannot revert to a viscous disk solution since strong flux advection continues to be maintained from the outer boundary.  Instead, we expect flux to continue to accumulate, working its way inward until the flow approaches a steady state with $\dot M \sim \dot M_0$ and MAD scaling all the way to the central object.  Note that the eventual state of the disc would be the true steady state, with $v_B \approx 0$, as opposed to the steady flux-advection cases we have been considering. 

This picture thus provides a possible scenario for creating a steady-state MAD.  However, it assumes that the toroidal field is able to self-regulate to $B_\phi \sim v_K \rho^{1/2} < B_z$.  There is no guarantee that this will happen, and we should consider the alternate possibility that the disc ``freezes'' near $r_0$, with so little turbulent activity at these high magnetization levels that $B_0$ is inadequate to drive accretion at even the relatively modest rate of $\dot M_0$.  In this case, we would expect the column density near $r_0$ to build up until it reaches a level where strong activity resumes, $B_\phi$ approaches $\sim 10 B_z$, and a period of MAD-like accretion with $\dot M \sim \dot M_{\rm mag}$ occurs.  Since the mass supply rate is smaller than $\dot M_{\rm mag}$, this would have to be episodic accretion with an on-off duty cycle of $\dot M_0/ \dot M_{\rm mag}$.        
 
\section{Discussion and conclusions}
\label{sec:conclusions}

We have presented a simple, analytic model for thin accretion discs containing a unidirectional vertical magnetic field, $B_z (r)$.  Because these discs are threaded by indestructible net magnetic flux, mean-field stochastic dynamo effects are not essential for maintaining the field and we ignore all of the commonly studied terms with the exception of an isotropic turbulent diffusivity, $\eta$, which is related to the turbulent viscosity via a turbulent magnetic Prandtl number, ${\rm Pr}_t$.  Importantly, we make no claims whatsoever as to the actual role of mean-field or other small-scale dynamo effects; they could well overwhelm the effects we discuss here.  However, if they do turn out to be relatively unimportant, our analysis shows that the large-scale field structure can be robustly governed by the balance between field amplification by shear and radial flow, versus turbulent diffusion and dissipation.

To isolate the basic features of the field evolution, we consider a local slab geometry with a scale height $H(r)$, with density $\rho(r)$ and $\eta(r)$ constant for $-H < z < H $ and zero elsewhere.  Assuming fixed angular velocity $\Omega (r)$ (taken to be Keplerian in our applications), we calculate the vertical dependences of $B_r$, $B_\phi$ and the radial velocity explicitly from the induction and  angular momentum conservation equations.  Both $B_r$ and $B_\phi$ exhibit sinusoidal behavior about the midplane, with $B_\phi$ showing additional linear behavior with $z$ due to the viscous stress and advection of radial magnetic flux.  

In addition to the simplification of a slab disc model, we neglect mass-loss from the disc.  Nevertheless, our model self-consistently estimates the angular momentum and energy extracted through the disc surface by magnetic stresses, and should be quantitatively consistent with disk-wind models in which mass loss is weak and the flow is very sub-Alfv\'enic near the disk surface.  One of the surface boundary conditions required by our model is $\xi = B_r(H)/B_z = \tan \theta$, where $\theta$ is the angle of a poloidal field line with respect to the disc axis. Our models encompass $\xi$ ranging from $\la H/r$ to $\sim O(1)$, and thus include cases where magnetocentrifugal winds are expected according to the argument of \cite{blandford82} ($\theta > 30^\circ$) as well as situations where winds are optional.  Yet all models lose significant angular momentum and energy through the disc surface, in addition to radial transport via viscous stress.  (We neglect radial transport via large-scale magnetic stress, which is generally small.) 

Our model does not self-consistently predict the value of $\xi$, which we provisionally associate with the global structure of the magnetic field (and any associated wind) above and below the disc.  For diffusive discs which accumulate little magnetic flux, we generally expect small values of $\xi$, but even for $\xi \sim O(H/r)$ the magnetic stresses contribute as much to driving accretion as the internal disc viscosity.  For $\xi \gg H/r$, and especially for the likely case $\xi \sim O(1)$, the large-scale stress vastly outweighs the viscosity and drives accretion at a speed roughly $r/H$ times the viscous speed, as pointed out by \cite{ferreira93}.  As also pointed out in this and other earlier work, the extraction of most angular momentum via vertical magnetic torques also implies that only a fraction $\sim H/r$ of the energy liberated by accretion is dissipated inside the disc.  One intriguing consequence of this is the absence of an extended radiation pressure-dominated regime in magnetically boosted accretion discs, even at relatively high $\dot M$.  

This effect of magnetically boosted accretion, by itself, does not resolve the flux accumulation problem for thin discs, identified by \cite{lubow94}. The ability of a disc to accumulate flux when $\xi \sim O(1)$ depends primarily on the shape of the poloidal field lines, which must possess sufficient convex-outward curvature near the disc surface.  In our simple model, this curvature is remarkably sensitive to a single parameter, $q = \sqrt{3} v_{{\rm A}z} H/ \eta $, with accumulation occurring for $q > \pi/2$.  Although the numerical values are likely to be different for more realistic models of disc structure, the reasoning behind these trends seems sufficiently robust for them to be taken seriously.  Arguing from numerical simulations in the literature and a speculative model for MRI saturation \citep{begelman23}, we guess that the value of $q$ is too small to trigger flux accumulation when the vertical field is weak compared to gas pressure ($\beta_z \ga 100$) but that this may switch at higher magnetizations.       

The sensitivity of the model behavior to $q$ underscores the importance of understanding the relationship between small-scale instabilities and large-scale magnetic structures.  Our lack of such understanding is most acute in the strongly magnetized cases.  For the level of turbulence we have adopted in this limit --- which assumes that the maximum pressure of the organized field is comparable to the turbulent energy density --- the squeezing effect of the organized field close to the midplane is of order unity, and is thus quantitatively but not qualitatively important. This begs the question of what happens if $B_\phi'$ reverses sign within the disc, or indeed how closely the level of turbulence is linked to the local behavior of the organized field.  Our closure seems reasonable on energetic grounds, since the same shear flow is responsible for both the turbulent and organized field.  But the argument from which the closure relation is derived \citep{begelman23} addresses a more specific context in which the turbulence in a highly magnetized disc is not driven directly by shear via MRI (which \citealt{begelman23} argues becomes too weak when the disc is highly magnetized), but rather by tearing instabilities driven by the vertical gradient of $B_\phi$, which in turn has been set up by the shear.  Since the strength of the instability depends on $B_\phi'$, might not the strength of the turbulence and thus $\eta$ be tied to the field structure locally, with dramatic variations as a function of height?   

These and related questions about highly magnetized accretion discs cannot be answered as yet, but it is important to ask them in the hope that they may guide future work.  In particular, it is hoped that upcoming simulations will be able to test some of the speculations here about the relationship between turbulent and organized magnetic fields in these systems.    

\subsection{Midplane outflows and field reversals}
\label{sec:elevated surface accretion}

The trigonometric behavior of field strengths and radial velocities with height in our model suggest that discs may exhibit internal radial and toroidal field reversals, accompanied by zones of outward radial flow.  Several simulations of magnetized discs have exhibited accretion dominated by regions offset from the midplane \citep{zhu18,lancova19,mishra20}, accompanied by weak inflow, or even outflow (``backflow'') near the equator. Analyzing the effect in their simulations, \cite{zhu18} argue that this accretion pattern is dominated by the $\phi z$ stress due to large-scale magnetic fields, as earlier noted by \cite{beckwith09}, rather than by viscosity operating on vertical shear \citep{kluzniak00,guilet12,guilet13}.
For our magnetically boosted models (i.e. neglecting internal viscosity), we find midplane outflow for $(2 n -1) \pi < q < 2 n \pi$, where $n \geq 1$ is an integer.  The corresponding disc exhibits $n$ internal field reversals. Such high values of $q$ would correspond to depressed levels of turbulent diffusivity compared to the fiducial scale $v_{{\rm A}z}H$; it remains to be seen under what conditions this would apply.  We note that the source of angular momentum (and energy) driving midplane outflow is the accretion occurring further from the midplane.  A portion of the angular momentum liberated in the inflow zone is deposited in the midplane region, driving outflow, while the remainder is carried away by magnetic stresses at the disc surface.  In all cases, the vertically integrated mass flux is inward and the layer nearest the disc surface is accreting.

The direction of flow can also be associated with the geometry of the poloidal field.  Changing the sign of $v_r(0)$ changes the sign of both $B_r$ and $B_\phi$ near the midplane.  Given our magnetic sign convention $B_z > 0$, equatorial outflow corresponds to $B_r <  0$, which means that the poloidal field lines projected in the azimuthal plane should bend inward away from the midplane.  This is clearly seen in Figure 3 of \cite{mishra20}, where the field lines bend inward in a region around the midplane characterized by weak outflow.  Moreover, our model predicts that $|B_r|$ should be maximized where $v_r$ changes sign and vanish in the middle of an elevated accretion zone --- both effects also apparent in \cite{mishra20} Figure~3.  This behavior gives the poloidal field lines the appearance of being pinched radially inward at some height above and below the midplane. Analogous information is displayed in Figure 15 of \cite{zhu18} where pinching is also seen, although the relationship between $B_r$ and $v_r$ is not so clear.  \cite{zhu18} Figure 5 also demonstrates the tendency of $B_r$ and $B_\phi$ to change sign at similar heights (our model actually predicts a small offset, due to the viscous contribution to inflow).     

\subsection{Global considerations and time-dependence}
\label{sec:global}

Unlike most previous analytic treatments of magnetized discs in the literature, we do not adopt radial self-similarity.  We exploit the fact that the generation of the toroidal field is generally faster than radial advection to obtain a quasi-local model.  This allows us to begin to address the time-dependent evolution of discs where matter and magnetic flux are injected externally. 

Our analysis strongly constrains the disc behaviors compatible with specified outer boundary conditions.  Under highly magnetized conditions and simple assumptions about disc turbulence, the accretion rate is a function of the magnetic field components, $\dot M \sim r B_\phi B_z/\Omega$, where $B_z$ is imposed and $B_\phi$ is determined by the balance between shear-amplification and vertical diffusion. For ``thinnish'' discs with $H/r \la$ a few tenths, we argue in \S\ref{sec:MRI} that $B_\phi \sim 10 B_z$, leading to the estimate of $\dot M_{\rm mag}$ in equation (\ref{eq:mdotmag}).  This is the maximum rate at which accretion can be driven by large-scale magnetic torques.  Thus, if an accretion flow is set up with a specified $B_0$ and $\dot M_0$ at radius $r_0$, and $\dot M_0$ exceeds the maximum magnetically boosted rate, then steady-state accretion can only occur through internal viscous stresses, i.e., as a Shakura-Sunyaev disc strongly dominated by gas pressure ($\beta_z \gg 1$), with 
 \begin{equation}
\label{eq:mdotSS4}
    \dot M = 5 \beta_z^{1/2}  {H\over r }  {r \over \Omega} B_z^2 > \dot M_{\rm mag} 
\end{equation}
and little flux advection.  

The situation becomes more complex if $\dot M_0$ is smaller than $\dot M_{\rm mag}$. For  
\begin{equation}
\label{eq:mdotSS5}
    (5-50) {H\over r }  {r_0 \over \Omega(r_0)} B_0^2  \la \dot M_0  < \dot M_{\rm mag} \ , 
\end{equation}
a Shakura-Sunyaev viscous disc should be a viable steady solution, provided that the mass injection process sets up a weakly magnetized system initially.  But a second, highly magnetized solution is also possible, leading either to a steady MAD-like disc with $\dot M = \dot M_0$ (if the turbulence self-regulates to just the right level, which seems unlikely) or an episodic, low-duty-cycle MAD with outbursts of $\dot M \approx \dot M_{\rm mag}$ interspersed with longer periods of quiescence.  Moreover, when $\dot M_0 \la  5 (H / r )  r_0 B_0^2/  \Omega(r_0) $, we surmise that the Shakura-Sunyaev solution is impossible, with an episodic MAD-like flow the only alternative.  

This unexpectedly rich phenomenology should be explored for its potential connection to state transition phenomena such as those in X-ray nova outbursts and changing-look AGN.  We plan to address this in subsequent work.

\section*{Acknowledgements}
I thank Phil Armitage, Jason Dexter, Geoffroy Lesur, Chris Reynolds, and Nico Scepi for valuable discussions during the preparation of this paper, and the anonymous referee for thoughtful and constructive criticism.  I acknowledge financial support from  NASA Astrophysics Theory Program grants NNX17AK55G and 80NSSC22K0826, and NSF Grant AST-1903335.   

\section*{Data Availability}
No new data was generated or analyzed to support the work in this paper.




\bibliographystyle{mnras}
\bibliography{biblio} 


\appendix

\section{Stability}
\label{sec:stability}

While a full stability analysis of our disc evolution equations is beyond the scope of this paper, we can examine the  quasi-local stability of our steady-state models (\S\ref{sec:steady}) by perturbing them with short-wavelength radial perturbations of the form  $y = y_0 (1 + \delta y)$ with $\delta y \propto e^{i(kr + \omega t)}$ and $H \ll k^{-1} \ll r$. The lower limit on $k^{-1}$ is intended to separate the scale of the perturbations from the presumed scale of velocity and magnetic fluctuations responsible for turbulent diffusivity.  We take $\delta\xi = 0 $, which is plausible since the external field configuration (and hence the inclination of the poloidal field at the disc surface) is probably governed by global external boundary conditions.  Defining $\nu \equiv (\omega/k) (H_0/\eta_0)$, we write the perturbed mass and flux equations in the form 
\begin{equation}
    \label{eq:masseq2}
    \nu (\delta \rho + \delta H) = \delta \rho + \delta H + \delta \bar v  \ , 
\end{equation}
\begin{equation}
    \label{eq:fluxeq2}
    \nu \delta B_z =  - \delta v_B   \ ,
\end{equation}
where $\delta v_B = (H_0/\eta_0) v_B$ since $v_B = 0$ in a steady state.  Using equations (\ref{eq:accretionspeed}) and (\ref{eq:fluxadv}) and exploiting the assumption that $H_0/r \ll 1$, we obtain, to the requisite order, 
\begin{equation}
    \label{eq:vperts}
    \delta v_B = - \left( {\pi \over 2} \right)^2  \delta q  \ ; \ \ \ \   \delta \bar v 
    = \left[ \left( {\pi \over 2} \right)^2 - 1 \right] \delta q + \delta v_{{A}z} \ , 
\end{equation}
where we have taken $\xi = {\rm Pr_t} = 1$ and used the definition of $q$, equation (\ref{eq:qdef}), to eliminate $\delta\eta$ in favor of the perturbed Alfv\'en speed.  The perturbed dynamical equations then become
\begin{equation}
    \label{eq:masseq3}
    (\nu - 1) (\delta \rho + \delta H) = \left[ \left( {\pi \over 2} \right)^2 - 1 \right] \delta q + \delta v_{{\rm A}z}   \ , 
\end{equation}
\begin{equation}
    \label{eq:fluxeq3}
    \nu \left(\delta v_{{\rm A}z} + {1\over 2} \delta \rho \right) =  \left( {\pi \over 2} \right)^2  \delta q   \ .
\end{equation}
Since these equations involve four perturbed quantities, we will need two additional conditions to obtain a dispersion relation.  These conditions depend on the specific equilibrium being perturbed.

\subsection{Weakly magnetized discs}
\label{sec:stabilityweak}

We first consider the weakly magnetized disc equilibrium discussed in \S\ref{sec:weaksteady}.  There we argue that $q$ depends on $\beta_z$, with higher levels of magnetization (lower $\beta_z$) corresponding to higher $q$. Therefore we adopt as the first auxiliary condition
\begin{equation}
    \label{eq:qcondweak}
    \delta q = -a \delta\beta_z = 2 a \left(\delta v_{{\rm A}z} - \delta H  \right)    \ ,
\end{equation}
where $a > 0$ is a constant and $\delta H$ serves as a proxy for $\delta c_s$ in this gas pressure-dominated case.  The value of $a$ cannot be determined without further insights into how MRI-driven turbulence saturates close to the steady-state threshold. 

The second condition comes from assuming that radiative equilibrium is maintained over perturbation timescales. The perturbed radiative flux is given by $\delta F_{\rm rad} = 14 \delta H - 2 \delta\rho$ for Kramers opacity and $\delta F_{\rm rad} = 7 \delta H -  \delta\rho$ for electron scattering. Expanding equation (\ref{eq:Feta3}) to the requisite order (which means retaining the $\cot q/q$ term) but ignoring the dissipation of the radial magnetic field, we obtain $\delta F_d = \delta\rho + 2 \delta H + \delta v_{{\rm A}z}$. Setting $\delta F_{\rm rad} = \delta F_d$ then allows us to express $\delta \rho$ in terms of $\delta H$ and $\delta v_{{\rm A}z}$:   
\begin{equation}
    \label{eq:deltarho}
    \delta \rho = 4 \delta H  - {1\over 3}\delta v_{{\rm A}z} \ {\rm (Kramers)}; \ \ \ \delta \rho = {5\over 2} \delta H  - {1\over 2}\delta v_{{\rm A}z} \ {\rm (e.s.)}   \ .
\end{equation}
Substituting into the evolution equations and computing the dispersion relation, we find that the case with Kramers opacity yields instability for $0.02 < a < 2.1$ while the electron scattering case is unstable for $0.02 < a < 2.7$; both are stable outside this range. We note that $a \ga 2$ represents a very steep dependence of $q$ on $\beta_z$, steeper than $\propto \beta_z^2$, and conjecture that the the value of $a$ is $\sim O(1)$ but somewhat smaller than 2.  This leads us to suggest that the weakly magnetized equilibrium is {\it unstable}.

\subsection{Strongly magnetized discs}
\label{sec:stabilitystrong}

At high magnetization, we expect to be in the regime where instability growth rates are increasing with $v_{{\rm A}\phi}$ \citep{das2018,begelman22,begelman23}. Since we have adopted the closure $v_{{\rm A}\phi} \propto H \propto v_{{\rm A}z}$ for this case, we have $\delta q = 2 \delta v_{{\rm A}\phi}- \delta \eta $. Since the maximum growth rate of instability in this limit is $\propto v_{{\rm A}\phi}$, the simple saturation argument presented in \cite{begelman23} suggests $\eta \propto v_{{\rm A}\phi}^3$, implying $q \propto v_{{\rm A}\phi}^{-1}$. We will therefore take $\delta q = - b \delta v_{{\rm A}\phi}$.  Making the necessary substitutions in the perturbed evolution equations, we find that the strongly magnetized equilibrium is unstable for $-1.63 < b < 0$; in other words, the expected behavior $b = 1$ implies {\it stability}.

A self-consistency condition on this stable regime leads to a lower limit on disc thickness.  For $c_s \la v_{{\rm A}\phi} \la (c_s v_k)^{1/2} $, the maximum MRI growth rate decreases with $v_{{\rm A}\phi}$ \citep{blaes94}, actually vanishing for $v_{{\rm A}\phi} = 2\sqrt{2} (c_s v_K)^{1/2}$ when $B_\phi \propto r^{-5/4}$ \citep{pessah05,das2018}. According to the saturation argument proposed by \cite{begelman23}, $\eta \propto v_{{\rm A}\phi}$ in this regime (provided that MRI is responsible for the turbulence), implying $b \sim -1$ and instability. In order to enter the stable regime of disc evolution, $H$ must satisfy
\begin{equation}
    \label{eq:pessahcond}
   {H\over r} \approx  {v_{{\rm A}\phi} \over v_K}  >  2 \sqrt{2} \left( {c_s \over v_K } \right)^{1/2}    \ .
\end{equation}
For $c_s/v_K$ in the typical range $\sim 10^{-3} - 10^{-2}$, this places a lower limit on $H/r$ of $\sim 0.1-0.3$, justifying our normalization of $H/r$ to 0.1 for the highly magnetized case.  Thus, a disc that enters the highly magnetized regime should evolve towards $H/r \ga O(0.1)$, which is close to the condition normally regarded as MAD \citep{begelman22}.  


\bsp	
\label{lastpage}
\end{document}